\documentclass[unnumsec,webpdf,modern,large]{mam-authoring-template}%

\usepackage{siunitx}  
\usepackage{placeins}
\usepackage{url}

\usepackage{hyperref}
\hypersetup{colorlinks,breaklinks,
            urlcolor=[rgb]{1,0,0},
            linkcolor=[rgb]{1,0,0},
            citecolor=[rgb]{1,0,0},
            filecolor=[rgb]{1,0,0}}
\usepackage{soul}
\usepackage{xcolor}

\graphicspath{{Fig/}}

\theoremstyle{thmstyleone}%
\theoremstyle{thmstyletwo}%
\theoremstyle{thmstylethree}%

\DeclareSIUnit\electron{\mathrm{e^-}}
\DeclareSIUnit\angstrom{\text{Å}}

\begin{document}

\journaltitle{Microscopy and Microanalysis}
\DOI{DOI HERE}
\copyrightyear{2025}
\pubyear{2025}
\access{Advance Access Publication Date: Day Month Year}
\appnotes{Original Article}

\firstpage{1}


\title[Clustering]{Unsupervised segmentation and clustering workflow for efficient processing of 4D-STEM and 5D-STEM data}

\author[1,$\ast$]{Serin Lee}
\author[2]{Stephanie M. Ribet}
\author[1]{Arthur R. C. McCray}
\author[3]{Andrew Barnum}
\author[1,$\ast$]{Jennifer A. Dionne}
\author[1,$\ast$]{Colin Ophus}

\authormark{Lee et al.}

\address[1]{\orgdiv{Department of Materials Science and Engineering}, \orgname{Stanford University}, \orgaddress{\street{Stanford}, \postcode{94305}, \state{CA}, \country{United States}}}
\address[2]{\orgdiv{National Center for Electron Microscopy}, \orgname{Molecular Foundry, Lawrence Berkeley National Laboratory}, \orgaddress{\street{Berkeley}, \postcode{94720}, \state{CA}, \country{United States}}}
\address[3]{\orgdiv{Stanford Nano Shared Facilities}, \orgname{Stanford University}, \orgaddress{\street{Stanford}, \postcode{94305}, \state{CA}, \country{United States}}}

\corresp[$\ast$]{Corresponding author. \href{email:email-id.com}{serinl@stanford.edu, jdionne@stanford.edu, cophus@stanford.edu}}

\received{Date}{0}{Year}
\revised{Date}{0}{Year}
\accepted{Date}{0}{Year}

\abstract{Four-dimensional scanning transmission electron microscopy (4D-STEM) enables mapping of diffraction information with nanometer-scale spatial resolution, offering detailed insight into local structure, orientation, and strain. However, as data dimensionality and sampling density increase, particularly for \textit{in situ} scanning diffraction experiments (5D-STEM), robust segmentation of structurally consistent behavior across sequential measurements becomes essential for efficient and physically meaningful analysis. Here, we introduce a clustering framework that identifies crystallographically distinct domains from 4D-STEM datasets. By using local diffraction-pattern similarity as a metric, the method extracts closed contours delineating spatially contiguous regions.
This approach produces cluster-averaged diffraction patterns that improve signal quality while reducing data volume by orders of magnitude, enabling rapid and accurate orientation, phase, and strain mapping. We demonstrate the applicability of this approach to \textit{in situ} liquid-cell 4D-STEM data of gold nanoparticle growth. Our method provides a scalable and generalizable route for spatially coherent segmentation, data compression, and quantitative structure–strain mapping across diverse 4D-STEM modalities. The full analysis code and example workflows are publicly available to support reproducibility and reuse. }


\maketitle

\section{Introduction}

Advances in four-dimensional scanning transmission electron microscopy (4D-STEM), building on earlier developments in scanning nanobeam diffraction and related techniques, now allow diffraction information to be mapped with nanometer resolution. \citep{ zuo2010scanning, eggeman2019scanning, ophus2019four}. Each probe position records a diffraction pattern, giving a spatially resolved view of local structure, strain, and orientation. Unlike selected-area diffraction, which can reach sub-micrometer spatial resolution depending on the aperture size but still averages diffraction signals over relatively large sample regions, and micro-diffraction using a focused probe, scanning nanobeam diffraction employs a probe that is orders of magnitude smaller, enabling nanoscale structural variations to be resolved. Previous studies have shown how automated crystal orientation mapping (ACOM), namely template matching of diffraction pattern libraries on 4D-STEM datasets allows crystal orientation mapping through open-source code frameworks~\citep{rauch2010automated, ophus2022automated, cautaerts2022free}. 

However, traditional analyses of the 4D-STEM datasets often rely on manually selected regions of interest or global thresholding to separate diffraction signals. These approaches can overlook subtle but physically meaningful variations across neighboring probe positions, especially in samples composed of closely spaced nanocrystals or partially coherent domains.\citep{mahr2021accurate,saha2025reuniting} Recent studies have employed unsupervised clustering methods, such as K-means clustering \citep{liu20244d,kim2025operando}, fuzzy c-means clustering \citep{martineau2019unsupervised}, non-negative matrix factorization \citep{eggeman2015scanning, uesugi2021non,allen2021fast,kimoto2025nonnegative}, linear iterative clustering \citep{guo2025picosecond}, density-based clustering using DBSCAN \citep{johnstone2020density}, and hierarchical clustering \citep{kimoto2024unsupervised} on 4D-STEM datasets, to identify regions with similar diffraction signatures. There are also approaches on segmenting and clustering 4D-STEM datasets through machine learning workflow, such as by using Gaussian mixture models \citep{liu2025end} and using variational autoencoders \citep{kho2025use}. While these techniques can reveal meaningful structural trends, they are sensitive to many user-specified parameters and therefore difficult to apply to varied datasets. As a result, they may struggle to capture irregularly shaped or hierarchically nested regions, which are common in polycrystalline or beam-sensitive systems. Moreover, while many of these methods are publicly available and compatible with open-source platforms, they are not always embedded within unified 4D and 5D-STEM specific workflows, which can introduce additional processing steps for end users.

\begin{figure*}[t]%
  \centering
  \includegraphics[width=\textwidth]{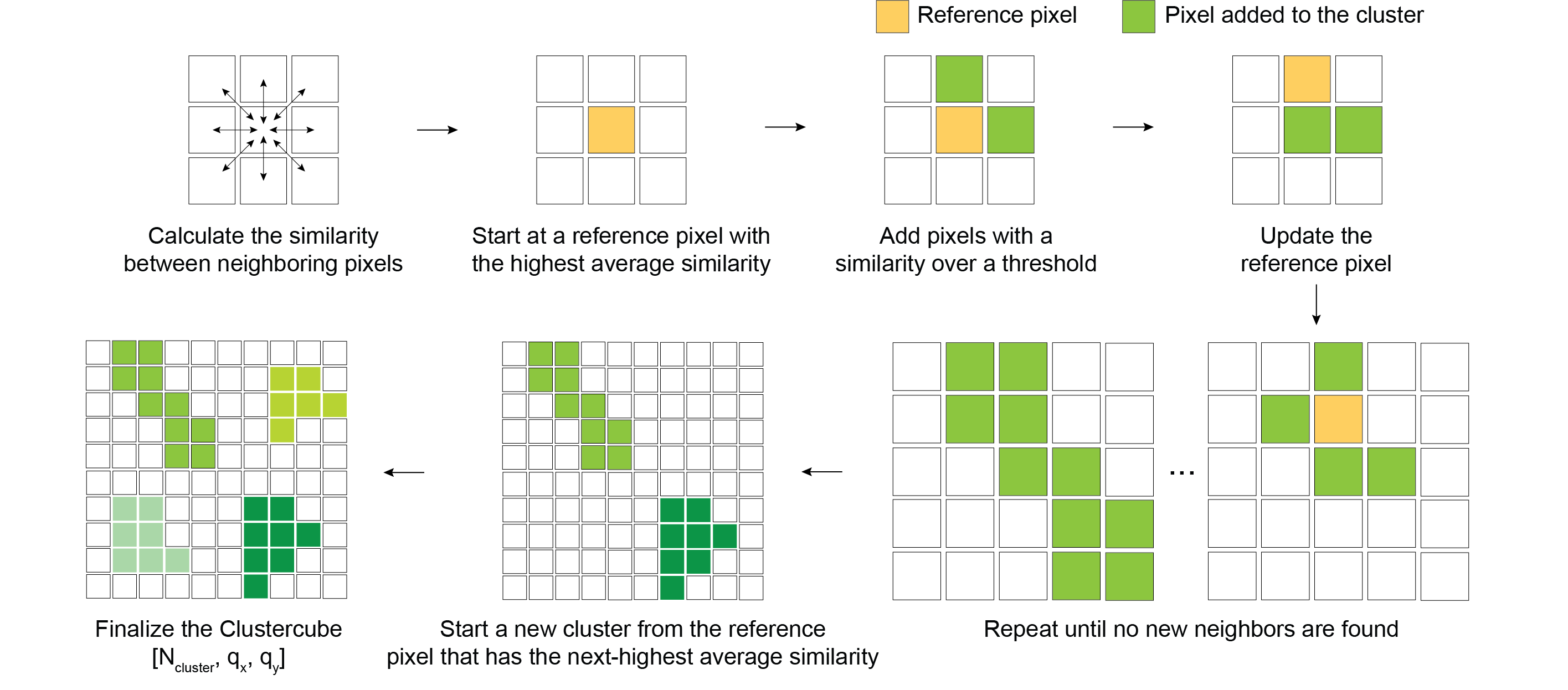}
  \caption{Schematic of the clustering process based on marching-square algorithm.}\label{fig1}
\end{figure*}

Modern detectors exacerbate these problems by enabling fast, high-resolution, low-dose measurements, conditions that often yield diffraction patterns with low signal-to-noise ratios. These innovations have heightened the need for robust and reliable clustering methods for interpreting the resulting high-dimensional data \citep{roccapriore2022automated}. 

Moreover, the massive size of 4D-STEM datasets demands computational methods that extract representative information while enabling efficient data reduction for interpretation, without compromising access to the underlying structural detail \citep{kimoto2024unsupervised}. This challenge becomes even more pronounced for \textit{in situ} 4D-STEM, i.e. 5D-STEM experiments \citep{miller2024results}, where datasets are accumulated over time, effectively adding a temporal dimension to the already large four-dimensional data structure \citep{pekin2019direct, miller2023continuous, da2022assessment, winkler2024texture, miller2025real, chen2025situ}. Importantly, such data reduction serves as an analytical abstraction rather than a replacement of the original dataset, preserving the ability to revisit the full measurements for detailed structural or boundary-specific analyses. Computationally efficient and physically meaningful interpretation is therefore indispensable for managing these large data volumes.

Furthermore, 5D-STEM experiments can often involve complex environments near the sample such as liquid or gas cells \citep{liu20244d, sun2024situ, miller2025situ, cheng2025direct, patrick2025situ}. The additional thickness associated with these experimental set-ups can degrade signal-to-noise ratios, making reliable segmentation and data reduction even more critical for quantitative analysis.

Here, we present a clustering framework based on the marching squares algorithm to automatically segment spatially coherent features within 4D-STEM datasets. By treating similarity maps between diffraction patterns as the metric to define clusters, our method extracts closed contours that trace the boundaries of crystallographically distinct regions. This approach provides a generalizable and computationally efficient means of identifying nanoscale domains, enabling researchers to analyze diffraction signals from regions that yield the most physically meaningful information. 
In addition to segmentation, the method reduces the real-space dimension from the total number of probes to the number of clusters(typically on the order of \(10^{-2}\) -- \(10^{-3}\)), producing a single representative diffraction pattern for each cluster. 
Averaging diffraction patterns within each cluster enhances the signal-to-noise ratio and facilitates downstream analysis such as orientation mapping, strain measurement, and structure determination. This combination of spatial segmentation, data reduction, and signal enhancement establishes a scalable workflow for interpreting large 4D-STEM datasets. In time-resolved experiments, the workflow can be applied sequentially to individual frames, making it well-suited for the analysis of large 5D-STEM datasets while the clustering itself is performed on the underlying 4D-STEM data.
 To ensure broad applicability, the framework is designed as a versatile tool with only a few user-tunable parameters, allowing easy adaptation to diverse datasets, and also implemented as a module in the open-source Python package \texttt{py4DSTEM}.

\section{Methods}\label{sec2}

\subsection{1. Preprocessing Using Correlative Pixel-based Filtering}\label{subsec1}
To enhance the local signal-to-noise ratio and suppress spurious noise before clustering, preprocessing using correlative pixel-based filtering procedure was applied in which each diffraction pattern was Gaussian-blurred directly in reciprocal space (i.e., along the $(q_x,q_y)$ coordinates of the diffraction pattern) using a method similar to that previously reported. \citep{chen2010bilateral, stangebye2025grain}. For each probe position $(x, y)$, the blurred diffraction pattern $I_{x,y}(q_x,q_y)$ 
was compared with its neighboring patterns within a predefined offset range 
$\{(\Delta x_i, \Delta y_i)\}$. 
This step produces a locally averaged dataset that preserves the spatial coherence 
of correlated features while minimizing uncorrelated background fluctuations.

A radial weighting function in reciprocal space $w(q_x,q_y)$ was applied in reciprocal space to reduce the influence of the central beam and emphasize higher-angle diffraction features relevant for structural discrimination, while suppressing detector edge artifacts.
The weight was constructed as a normalized radial function:

\[
w(q_x,q_y) = 
\left[
\frac{\sqrt{(q_x - \bar{q}_x)^2 + (q_y - \bar{q}_y)^2}}
{\sqrt{N_{q_x} N_{q_y}}}
\right]^2
\]

where $(\bar{q}_x, \bar{q}_y)$ denote the mean in the reciprocal space, 
and $N_{q_x}, N_{q_y}$ are the diffraction pattern dimensions. 

Each diffraction pattern was then weighted and normalized. 
The weighted mean intensity was subtracted, and the pattern was normalized 
by its L2 norm:

\[
\tilde{I}_{x,y} = 
\frac{
\left(I_{x,y} - \frac{\sum I_{x,y} w}{\sum w}\right) w
}{
\sqrt{\sum \tilde{I}_{x,y}^2}
}
\]

The correlation coefficient ($R_i$) between the reference pixel $(x,y)$ 
and each neighboring pixel $(x+\Delta x_i, y+\Delta y_i)$ was computed as:

\[
R_i(x,y) = 
\sum_{q_x,q_y} 
\tilde{I}_{x,y}(q_x,q_y)\,
\tilde{I}_{x+\Delta x_i,\,y+\Delta y_i}(q_x,q_y)
\]

The neighbor offsets $(\Delta x_i, \Delta y_i)$ were defined within a circular 
footprint of radius~$r = 4$, corresponding to 81 neighboring probe positions.

The correlation values were mapped into a normalized range $[0,1]$ 
using a predefined correlation range $(R_{\min}, R_{\max}) = (0.85, 0.95)$:

\[
\hat{R}_i = 
\mathrm{clip}\left(
\frac{R_i - R_{\min}}{R_{\max} - R_{\min}},\, 0,\, 1
\right)
\]

Finally, the intensity at each probe position was 
refined by weighted accumulation of its neighboring patterns 
according to their correlation strength:

\[
I'_{x,y} = 
\frac{I_{x,y} + 
\sum_i \hat{R}_i\, I_{x+\Delta x_i,\,y+\Delta y_i}}
{\sum_i \hat{R}_i}
\]

This filtering process effectively averages each diffraction pattern 
with its correlated neighbors while preventing over-smoothing across distinct regions. The resulting filtered dataset was subsequently used 
for similarity calculation and marching-square clustering. This preprocessing step includes adjustable parameters, including the neighborhood offset and the correlation range, which may be tuned depending on the specimen contrast and noise characteristics of the dataset. A step-by-step implementation and guidance on parameter selection are provided in the accompanying \texttt{\detokenize{py4DSTEM_tutorials}}  repository.

\subsection{2. Clustering of 4D-STEM Data}\label{subsec1}
To identify spatially coherent regions exhibiting similar diffraction behavior, we developed a marching-square-based clustering algorithm that segments 4D-STEM data based on local correlation of diffraction patterns. The algorithm operates in three stages: (a) computation of a pixel-wise similarity matrix, (b) real-space background masking and thresholding, (c) clustering using a marching-square algorithm, and (d) refining cluster and averaging diffraction cubes (Fig. ~\ref{fig1}).

\subsection{A. Similarity Matrix Calculation}

Each 4D-STEM dataset $D(r_x, r_y, q_x, q_y)$ is treated as a two-dimensional array of diffraction patterns indexed by probe positions $(x, y)$.  
For each probe position, the diffraction pattern $I_{x,y}(q_x,q_y)$ is optionally Gaussian-blurred with a kernel width of $\sigma = 3$ to reduce high-frequency noise (an additional correlative filtering step is applied and discussed later).  

To restrict the similarity calculation to a desired reciprocal-space region, a diffraction-space mask $M_q(q_x, q_y)$ can be applied:

\[
I'_{x,y}(q_x,q_y) = I_{x,y}(q_x,q_y) \cdot M_q(q_x,q_y)
\]

 $M_q(q_x,q_y)$ is defined as an annular mask in reciprocal space, excluding both the central direct-beam region and the highest-angle scattering beyond a selected radial cutoff. This geometry suppresses contributions from the intense central spot and edge artifacts while retaining intermediate scattering features that contain the most structurally informative diffraction signals.

The similarity ($S(x, y, \Delta)$)  between each diffraction pattern and its neighbors is computed using a normalized cosine correlation:

\[
S(x, y, \Delta) =
\frac{\sum_{q_x,q_y} I'_{x,y}(q_x,q_y) \, I'_{x+\Delta_x,\, y+\Delta_y}(q_x,q_y)}
{\sqrt{\sum_{q_x,q_y} I'_{x,y}(q_x,q_y)^2} \;
 \sqrt{\sum_{q_x,q_y} I'_{x+\Delta_x,\, y+\Delta_y}(q_x,q_y)^2}}
\]

Here,\[
\Delta = 
\{(-1,-1),(-1,0),(-1,1), (0,-1), (0,1), 
(1,-1), (1,0), (1,1)\}
\]represents the eight fixed nearest-neighbor offsets used in the marching-square scheme.  
The resulting 3D similarity array $S(x, y, n)$ stores one similarity value per neighbor direction $n$. For edge pixels, the similarity calculation includes only the available nearest neighbors, so the neighborhood is effectively truncated at the dataset boundaries while the similarity metric is computed in the same manner as for interior(non-edge) pixels.

\begin{figure*}[!t]%
  \centering
  \includegraphics[width=\textwidth]{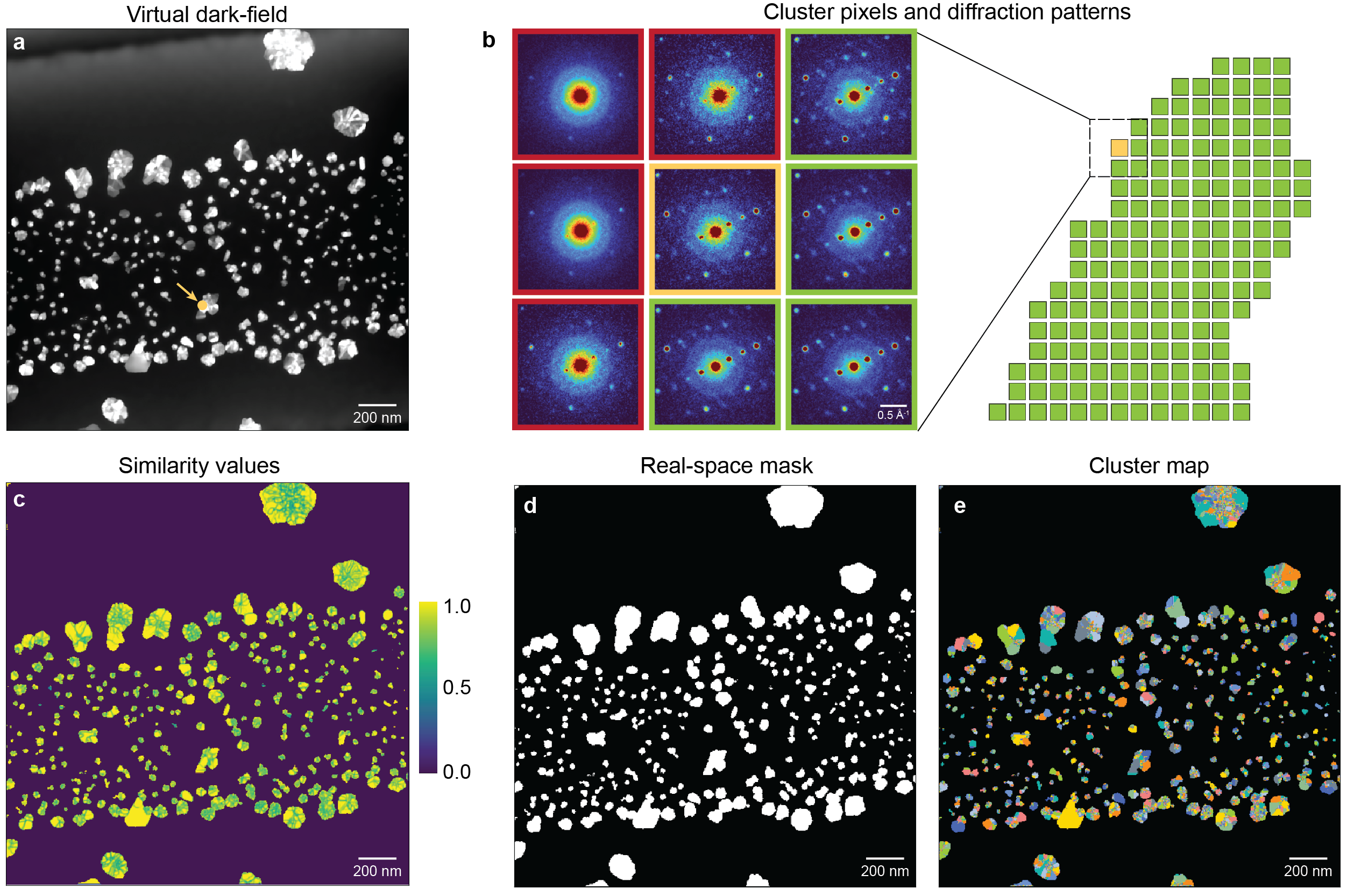}
  \caption{Applying clustering to the preprocessed 4D-STEM dataset. a. Virtual dark field image of the Au nanoparticles formed by the electron-beam induced reduction of Au cation precursors. b. (left) Diffraction pattern from the reference pixel (yellow box; marked as the yellow dot and the yellow arrow in panel a). Neighboring pixels with diffraction patterns above the similarity threshold are highlighted in green, while those below the threshold are shown in red. (right) Schematic illustration of the resulting cluster, where each square corresponds to a probe position within the cluster. c. Plot of the similarity values ($\bar{S}(x,y)$), where the color range spans from 0 to 1. d. Real-space mask generated for the background thresholding. e. Cluster map, where colors were iterated over ten distinct color codes to differentiate clusters.  }\label{fig2}
\end{figure*}

\subsection{B. Real-Space Masking and Background Thresholding}

To suppress background regions with low signal or outside the area of interest and reduce computational cost, the similarity values are first averaged over all neighbor directions to obtain a scalar map:
\[
\bar{S}(x,y) = \frac{1}{N} \sum_{n=1}^{N} S(x,y,n)
\]

where $N = 8$ represents the number of neighbor directions. 
A real-space mask $M_r(x,y)$ is then applied to this averaged similarity map to exclude non-sample or low-signal regions:

\[
\bar{S}'(x,y) = \bar{S}(x,y) \cdot M_r(x,y)
\]

Pixels outside the mask (i.e. $M_r = 0$) are excluded from further analysis.

\subsection{C. Marching-Square Clustering}

Clustering proceeds by iteratively growing connected regions whose local similarity exceeds a predefined threshold.  
The process begins from the unassigned pixel with the highest $\bar{S}(x,y)$, which serves as the seed of a new cluster.  
From this seed, the algorithm inspects its eight neighboring pixels defined by $\Delta$.  
A neighbor $(x+\Delta_x,\, y+\Delta_y)$ is added to the cluster if

\[
S'(x, y, \Delta) \geq T
\]

where $T$ is the user-defined similarity threshold, and if the neighbor has not already been assigned to another cluster.  
Each accepted pixel becomes a new node for expansion, forming a recursive marching loop until no new neighboring pixels satisfy the threshold.  
When cluster growth halts, the current region is finalized, and the next unassigned pixel with the highest remaining $\bar{S}(x,y)$ is used to seed a new cluster.  
This process repeats until all pixels are either assigned to a cluster or masked out.

\subsection{D. Cluster Refinement and Averaged Diffraction Cubes}

Each resulting cluster $C_i$ contains the real-space coordinates of its member pixels, where $i$ refers to cluster number.  
To suppress noise-induced fragments, clusters smaller than a specified minimum size can be excluded.  
For each remaining cluster, an averaged diffraction pattern is generated:

\[
 I_{C_i}  =
\frac{1}{|N_i|} \sum_{(x,y)\in C_i} I_{x,y}(q_x,q_y)
\]

Here $N_i$ indicates the number of pixels contained in the cluster $C_i$. This produces a reduced cluster-averaged diffraction cube representing the mean diffraction behavior of each region, which now reduces the data from  $D(x, y, q_x, q_y)$ to a 3D representation of $D(N_{\text{cluster}}, q_x, q_y)$.

\begin{figure*}[!t]%
  \centering
  \includegraphics[width=\textwidth]{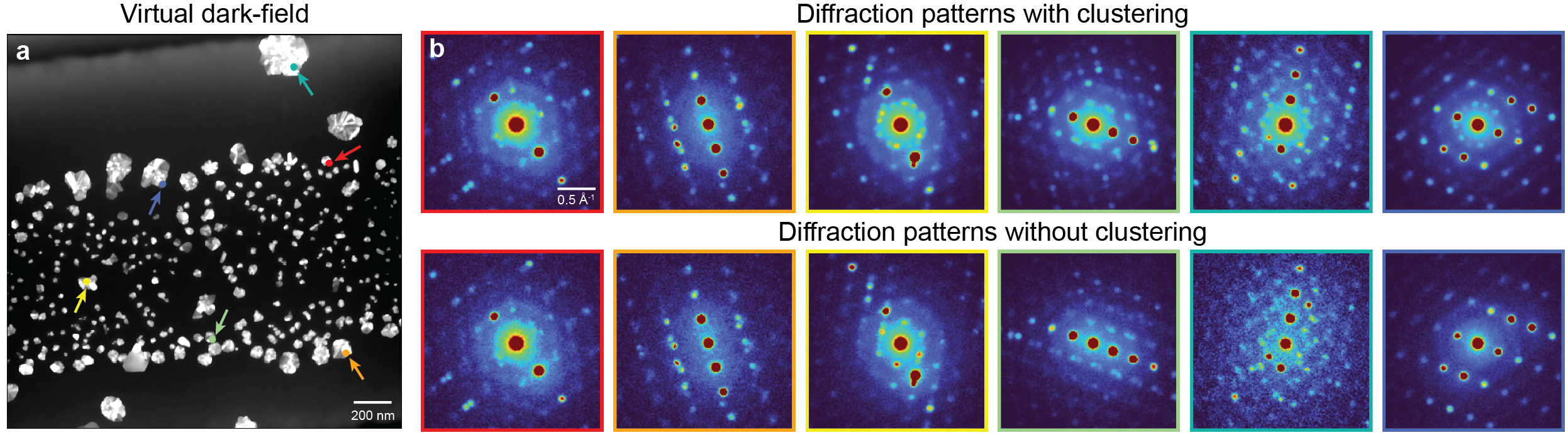}
  \caption{
    Comparing diffraction patterns with and without clustering.
    (a) Virtual dark-field image with colored markers indicating the six cluster seeds.
    (b) Diffraction patterns from the same positions with clustering (top) and without clustering (bottom).
  }\label{fig3}
\end{figure*}

\subsection{3. 4D-STEM experiment}\label{subsec1}

As a model system to demonstrate the clustering of the 4D-STEM dataset, we used Au nanoparticles grown by the electron-beam induced radiolysis in liquid cell TEM. Under electron irradiation, radiolysis of water produces reducing species such as hydrated electrons and radicals, which reduce the dissolved Au precursor to form nuclei that subsequently grow into nanoparticles. In this system, the nanoparticles nucleate heterogeneously on the SiNx membrane of the liquid-cell chip. \citep{woehl2012direct,schneider2014electron, woehl2020electron,lee2023temperature,fritsch2025influence,lee2024temperature}. We introduced aqueous 1mM HAuCl$_{4}$ solution into a liquid-cell holder (Insight Chips, Flow-holder) with a 80 nm-thick channel (Insight Chips, Nano Channel Chips), and the Au nanoparticles were grown under the electron beam irradiation with an electron flux of \qty{1.14}{\electron \per \angstrom \squared\per \second}. The left panel of the Fig.~\ref{fig2}a shows the virtual dark-field image of the Au nanocrystals formed after a total 45 s of irradiation. The 4D-STEM data was collected on a Thermo-Fisher probe-corrected Spectra microscope. The 4D-STEM time series data were collected using a Dectris Arina detector. The accelerating voltage was 300 kV. The scan consisted of a  $512 \times 512$ array of real-space probe positions, and at each probe position, a diffraction pattern of $192 \times 192$  detector pixels was recorded, with a convergence semi-angle of 0.622 mrad, and a dwell time of 50 $\mu$s per probe position, giving a total effective frame time of 13.1 s. ~\citep{stroppa2023stem} The analysis presented here corresponds to a representative single time frame acquired while liquid remained present in the field of view, and the extension of this workflow to fully time-resolved 5D-STEM datasets represents an application rather than a result explicitly demonstrated in this study.

The yellow-boxed image of the right panel of the Fig.~\ref{fig2}a shows the reference pixel of the first cluster, and the green-boxed images show the neighboring pixels that were added to the cluster since they had a similarity higher than the threshold (0.9 in the example). Fig.~\ref{fig2}b shows the preprocessed diffraction patterns of the first 30 pixels that were added to the cluster during the iterative process. 
To apply the marching-square clustering, first the $S(x,y)$ was calculated (Fig.~\ref{fig2}c) using the preprocessed 4D-STEM dataset. The background was thresholded using the triangle method \citep{van2014scikit} to produce the real-space mask (Fig.~\ref{fig2}d). We then applied our clustering algorithm, and the resulting clusters are  shown in Fig.~\ref{fig2}e. 

\subsection{4. Orientation and Strain Mapping}\label{subsec1}

The clustered datacube was used to calculate orientation and strain maps using the workflows with ACOM \citep{ophus2022automated} from \texttt{py4DSTEM} package \citep{savitzky2021py4dstem}. Note that depending on the experiment, preprocessing steps such as descan correction may be applied to improve the matching, and scan distortions and diffraction calibration should be verified prior to quantitative analyses such as template matching or strain mapping.

\section{Results}\label{sec3}

We first compare diffraction patterns from the same probe positions before and after applying the clustering algorithm. In the virtual dark-field image (Fig.~\ref{fig3}a), the colored markers and the arrows indicate the seed pixels corresponding to the six identified clusters. These seeds are selected sequentially during the clustering process, but are shown together in the figure for visualization. The corresponding averaged diffraction patterns for each cluster are displayed in the top row of Fig.~\ref{fig3}b. For comparison, the bottom row shows the diffraction patterns at the same seed pixel positions without to clustering. Note that both rows in Fig.~\ref{fig3}b show diffraction patterns after preprocessing of the 4D-STEM data using the correlative pixel-based filtering as described above.  Fig.~\ref{figS1} shows the diffraction patterns without preprocessing and clustering. The clustered averages exhibit an improved signal-to-noise ratio, enhancing the visibility of diffraction features, particularly at higher scattering angles, and thereby demonstrating the effectiveness of clustering in recovering weak high-angle signals. To further illustrate the fine-scale clusters within larger grains, representative cluster-averaged diffraction patterns from multiple regions of a single nanoparticle are shown in  Fig.~\ref{figS2} , demonstrating distinct and coherent crystallographic variations across neighboring clusters.

\begin{figure*}[!t]%
  \centering
  \includegraphics[width=\textwidth]{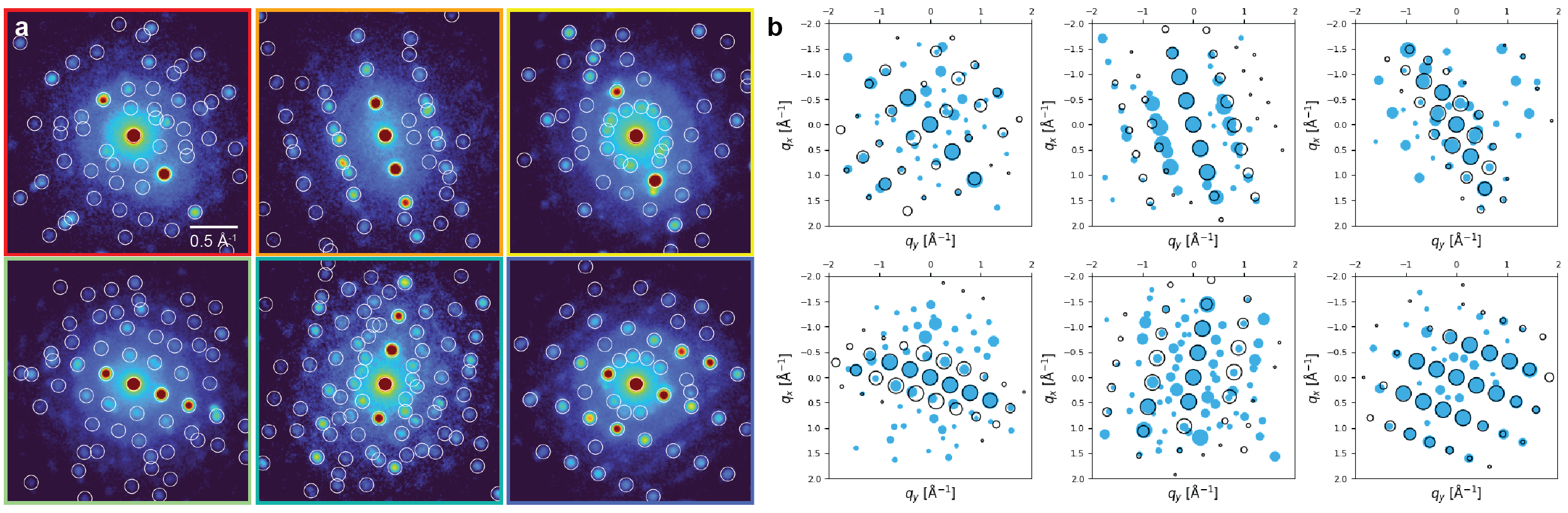}
  \caption{Bragg disk detection with clustering. Bragg disk detector (a) and orientation mapping by ACOM template matching (b) on diffraction patterns. Blue spots represent the experimentally detected Bragg disks, while black circles indicate the predicted disk positions from the best-matching simulated template used for ACOM indexing.}\label{fig4}
\end{figure*}

We next perform the ACOM analysis on the clustered dataset. Fig.~\ref{fig4}a presents representative clustered diffraction patterns overlaid with detected Bragg disks used for parameter tuning, while Fig.~\ref {fig4}b shows examples of ACOM template matches. 
The clustering substantially improves computational efficiency in the Bragg disk matching step. Without clustering, matching must be performed for every probe position, corresponding to \( N_\text{probe} = 512 \times 512 = 2.62 \times 10^5 \) diffraction patterns. With clustering, matching is instead carried out once per cluster, reducing the total operations to \( N_\text{cluster} \). The resulting computational cost scales as
\[
R = \frac{N_\text{cluster}}{N_\text{probe}},
\]
where \( R \ll 1 \), typically on the order of \(10^{-2}\) -- \(10^{-3}\) depending on the dataset. This reduction by several orders of magnitude enables rapid and scalable orientation mapping while maintaining accuracy in the ACOM results, which will be discussed in detail below by comparing the cross-validation angular errors. Moreover, compared to the ACOM analysis on dataset without clustering (Fig.~\ref{figS3}a,b) and the raw dataset without clustering and preprocessing (Fig.~\ref{figS3}c,d), clustering detects higher scattering angle Bragg disks along with less overlapping. We note that the demonstrated Au nanoparticle dataset exhibits strong diffraction contrast between crystalline particles and the surrounding liquid matrix, which facilitates segmentation and contributes to the observed reduction efficiency. The reduction factor
$R$ therefore depends on the degree of structural heterogeneity and contrast in a given specimen. Although certain parameters (e.g., similarity thresholds and reciprocal-space masks) remain user-defined, the underlying segmentation framework is general and can be adapted to systems with more subtle structural variations.

\begin{figure*}[!t]%
  \centering
  \includegraphics[width=\textwidth]{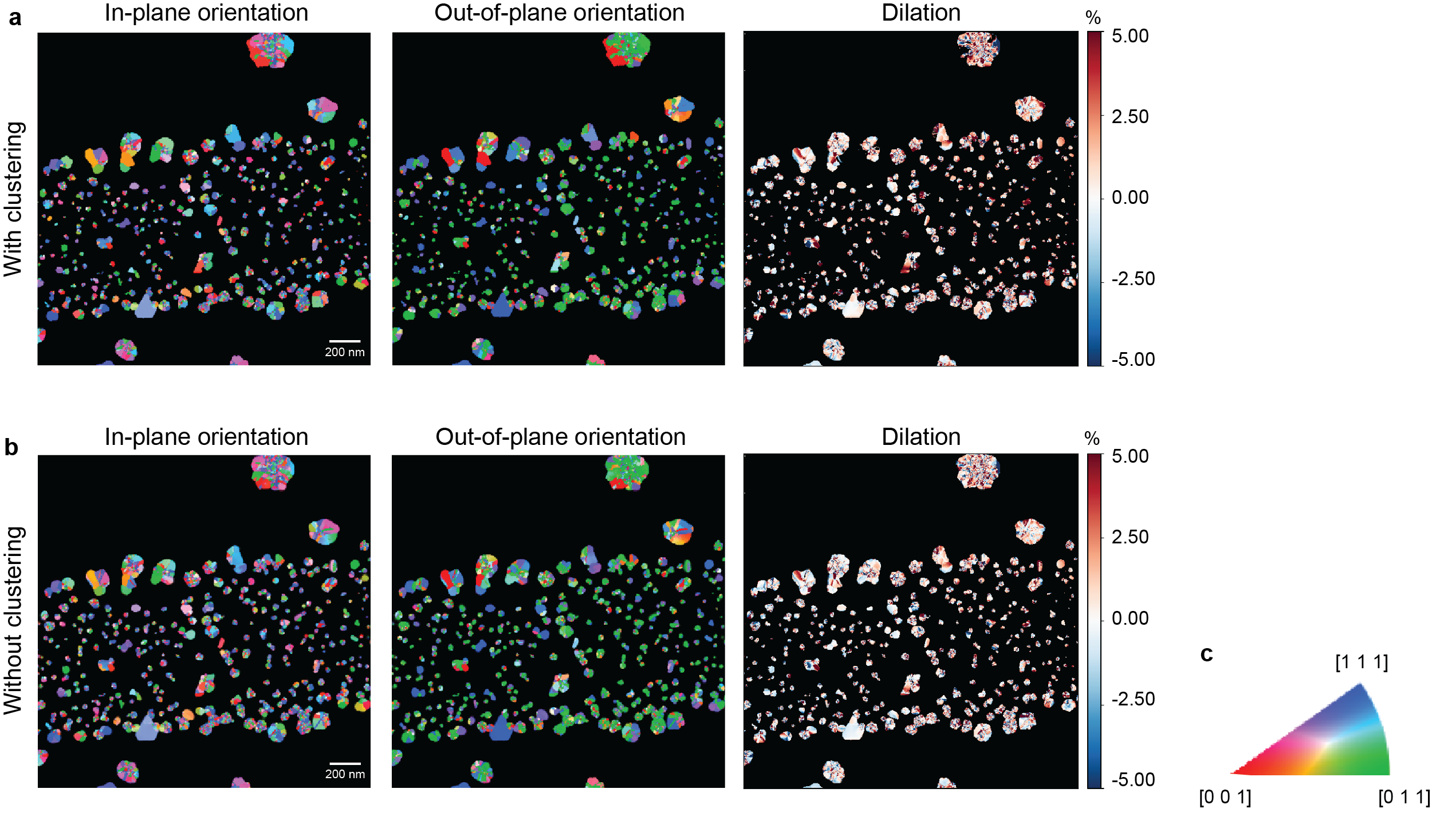}
  \caption{Orientation and strain mapping with (a) and without clustering (b). a. (left) In-plane orientation map, (middle) out-of-plane orientation map, and (right) dilation map of a clustered dataset.  b. (left) In-plane orientation map, (middle) out-of-plane orientation map, and (right) dilation map of a non-clustered dataset. c. Color legend of the orientation maps.  }\label{fig5}
\end{figure*}

We subsequently perform orientation and strain mapping, as summarized in Figure~\ref{fig5}. 
Both the in-plane (left, Fig.~\ref{fig5}a) and out-of-plane (middle, Fig.~\ref{fig5}a) orientations are determined using the conventional ACOM workflow, with the mapping performed by assigning to each real-space position the orientation of its corresponding cluster. The orientations in Fig.~\ref{fig5}a are represented as in-plane and out-of-plane components relative to the substrate normal to highlight possible growth-induced texture. Because the Au nanoparticles nucleate and grow on the SiN membrane of the liquid-cell chip, a substrate-induced orientation bias toward dense crystallographic directions (i.e. [111] and [110]) is physically plausible. To quantitatively evaluate the robustness of the ACOM results after preprocessing and clustering, we performed a checkerboard cross-validation on the 4D-STEM dataset. The scan positions of raw, preprocessed, and preprocessed and clustered datasets were each divided into two complementary 50/50 subsets, and orientation maps were independently reconstructed from each subset. For each pixel, the orientation from one map was compared with the four neighboring pixels in the complementary map, and the minimum angular deviation was assigned as the cross-validation error. The mean of this error map provides a measure of the consistency of the orientation solution across independently sampled diffraction data. For the raw 4D-STEM data without preprocessing or clustering, the mean angular error was 7.32°, while preprocessing alone reduced the error to 5.19°, and preprocessing combined with clustering yielded a mean error of 2.03°. These results suggest the improved robustness of the orientation determination after clustering. The resulting cross-validation angular error maps are provided as Fig.~\ref{figS4} in the Supporting Information.

For strain analysis, we introduce a physically consistent approach that leverages both clustered and non-clustered Bragg peak information  (right, Fig.~\ref{fig5}a). From ACOM maps, the strain at each position can be calculated by comparing to the reference \texttt{cif} file as described in ~\citep{ribet2025multi}. 
To apply this to clustered data, the orientation map obtained after clustering is used as a reference lattice orientation, while the Bragg peak positions extracted from the non-clustered diffraction patterns are used to quantify local lattice distortions. This approach leverages the improved orientation stability obtained through clustering while preserving the spatial resolution of the raw diffraction measurements. However, when compared to the dilation map resulting from the non-clustered data (right, Fig.~\ref{fig5}b), in some probe positions, reliable strain values cannot be obtained because the local diffraction pattern contains contributions from more than one crystalline domain. When diffraction signals from neighboring or overlapping grains are superimposed within the probe volume, the resulting mixed diffraction pattern cannot be uniquely associated with a single reference lattice, preventing a robust strain measurement. Strategies to mitigate these limitations, including both experimental approaches and computational refinements, will be discussed later.

\section{Discussion}\label{sec4}

The marching-square-based clustering framework offers a fast and generalizable route for identifying spatially coherent regions in 4D-STEM data. While it reduces computational time, it can also improve the accuracy of the template matching at higher scattering angles by mitigating the effects of local mistilt and mosaicity within a grain, an averaging effect conceptually similar to that exploited in precession electron diffraction methods. \citep{vincent1994double, midgley2015precession}. However, it also has limitations to be resolved. The algorithm assumes that each probe position is dominated by a well-defined diffraction pattern representative of a unique local structure. While cluster averaging can enhance the visibility of higher-angle diffraction features, clusters may occasionally include pixels from slightly different crystalline domains, which can lead to superposition of diffraction features depending on the dataset and clustering parameters. \citep{martineau2019unsupervised} These mixed diffraction conditions also limit strain extraction, since reliable strain measurements require that experimental Bragg peaks can be uniquely associated with a single reference lattice. Clustering approaches are therefore most effective when the dataset contains a significant fraction of pixels with relatively unmixed diffraction signals. In strongly polycrystalline systems, multiple crystalline domains may contribute simultaneously to the diffraction pattern at a given probe position, for example due to overlapping grains, twin boundaries, or vertically stacked grains within the probe size. In such cases the local diffraction signals become mixed, which can make similarity-based segmentation less distinct and lead to ambiguous similarity metrics and less stable cluster boundaries.

In these cases, experimental approaches such as precession \citep{vincent1994double, midgley2015precession} or multi-angle beam-tilt acquisition \citep{ribet2025multi} can help reduce overlap effects. Computational strategies such as hierarchical clustering or machine-learning assisted clustering may further refine segmentation of complex microstructures. In the current implementation, clusters are seeded sequentially from pixels with the highest similarity values using a single similarity threshold for efficiency. However, more complex microstructures, such as boundaries between domains sharing common reflections(e.g. twin boundaries), may benefit from parameter exploration or approaches that evaluate similarity threshold at multiple levels. These developments could better resolve structural relationships such as twins, grain boundaries, or other crystallographic variants.

Despite these limitations, the proposed method offers significant practical advantages for processing 4D-STEM datasets. Implemented as an open-source user-friendly module within the \texttt{py4DSTEM} ecosystem, the algorithm requires only a few tunable parameters, primarily the similarity threshold and minimum cluster size, making it broadly accessible to users with varying computational backgrounds. A major advantage of this framework is the substantial reduction in effective data size, which greatly facilitates data handling, visualization, and storage. This benefit becomes even more pronounced in \textit{in situ} 4D-STEM (or 5D-STEM) experiments, where continuous acquisition rapidly produces massive datasets. Moreover, the low computational cost enables efficient processing even on standard CPUs, eliminating the need for specialized hardware. Together, these features establish a scalable and practical workflow for multidimensional electron microscopy analysis.

\section{Conclusion}

We have developed a clustering algorithm for spatially coherent segmentation of 4D-STEM datasets. By defining local similarity between diffraction patterns and recursively tracing connected regions above a threshold, the method isolates nanoscale domains that exhibit consistent crystallographic behavior. This clustering enhances diffraction signal quality through regional averaging and reduces the computational load of subsequent analyses such as phase, orientation, and strain mapping by a significant amount, up to several orders of magnitude. The resulting cluster-averaged diffraction cubes enable efficient and physically interpretable quantification of orientation and strain, bridging the gap between raw multidimensional datasets and meaningful structural descriptors. Applied to \textit{in situ} 4D-STEM of Au nanocrystals grown under electron irradiation, the approach robustly delineates grain boundaries and strain gradients even in noisy liquid-cell environments. More broadly, this framework offers a scalable foundation for real-time segmentation, data reduction, and correlation analysis in next-generation multidimensional electron microscopy.


\section{Competing interests}
No competing interest is declared.

\section{Data availability}
The 4D-STEM dataset of the Au nanoparticle in the liquid cell TEM is available online. (\url{https://zenodo.org/records/18167694}) The analysis code is part of \texttt{py4DSTEM}
(\url{https://github.com/py4dstem/py4DSTEM.git}), and the example Jupyter Notebook is in
\texttt{\detokenize{py4DSTEM_tutorials}} repository (\url{https://github.com/py4dstem/py4DSTEM_tutorials}). A preprint was posted on \texttt{\detokenize{arXiv}} (\url{https://arxiv.org/abs/2601.17262}) \citep{lee2026unsupervised}.

\section{Acknowledgments}

S.L. and J.A.D. acknowledge financial support from the Office of Basic Energy Sciences, U.S. Department of Energy, Division of Materials Science and Engineering, DE-AC02-76SF00515. Additionally, S.L. and J.A.D acknowledge the financial support from the U.S. Department of Energy, Office of Science, National Quantum Information Science Research Centers as part of the Q-NEXT center. Work at the Molecular Foundry was supported by the Office of Science, Office of Basic Energy Sciences, of the U.S. Department of Energy under Contract No. DE-AC02-05CH11231. The authors thank Professor Josh Kacher for helpful suggestions on the noisy diffraction pattern analysis. The authors thank Dr. Parivash Moradifar for the helpful discussions related to this work.

\bibliographystyle{unsrtnat}
\bibliography{reference}

\onecolumn

\renewcommand{\thefigure}{S\arabic{figure}}
\setcounter{figure}{0}

\section*{Supplementary Material}

\noindent{\Large \textbf{Unsupervised segmentation and clustering workflow for efficient processing of 4D-STEM and 5D-STEM data}}

\vspace{1em}

\noindent{\large Serin Lee$^{1,\ast}$, Stephanie M. Ribet$^{2}$, Arthur R. C. McCray$^{1}$,
Andrew Barnum$^{3}$, Jennifer A. Dionne$^{1,\ast}$, Colin Ophus$^{1,\ast}$}

\vspace{0.8em}

\noindent{\footnotesize $^{1}$Department of Materials Science and Engineering, Stanford University, Stanford, CA 94305, United States\\
$^{2}$National Center for Electron Microscopy, Molecular Foundry, Lawrence Berkeley National Laboratory, Berkeley, CA 94720, United States\\
$^{3}$Stanford Nano Shared Facilities, Stanford University, Stanford, CA 94305, United States\\[0.6em]
$\ast$Corresponding authors: \text{serinl@stanford.edu}, \text{jdionne@stanford.edu}, \text{cophus@stanford.edu}}

\clearpage

\FloatBarrier
\begin{figure}[!t]
  \centering
  \includegraphics[width=\linewidth]{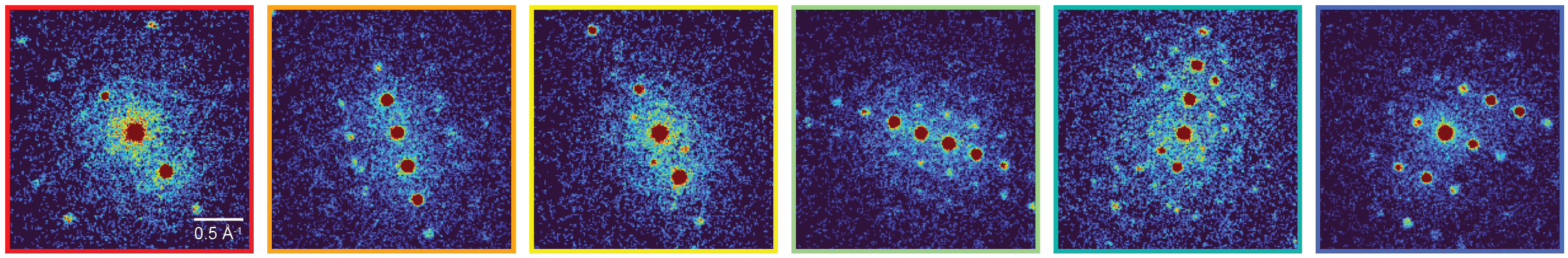}
  \caption{Diffraction patterns of raw data without preprocessing and clustering. The color codes correspond to (Fig.~\ref{fig3})}
  \label{figS1}
\end{figure}

\clearpage

\FloatBarrier
\begin{figure}[!t]
  \centering
  \includegraphics[width=\linewidth]{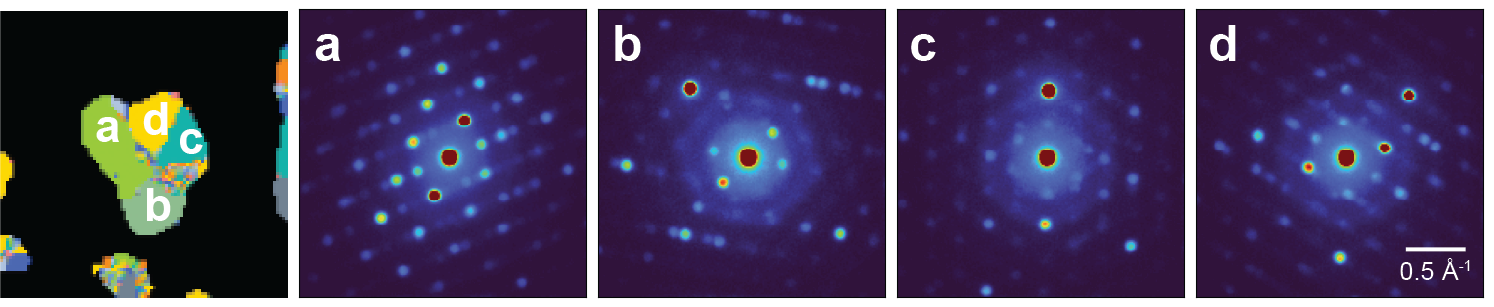}
  \caption{Example of cluster-averaged diffraction patterns from multiple regions of a single nanoparticle shown in  (Fig.~\ref{fig2}e)}
  \label{figS2}
\end{figure}

\clearpage

\FloatBarrier
\begin{figure}[!t]
  \centering
  \includegraphics[width=\linewidth]{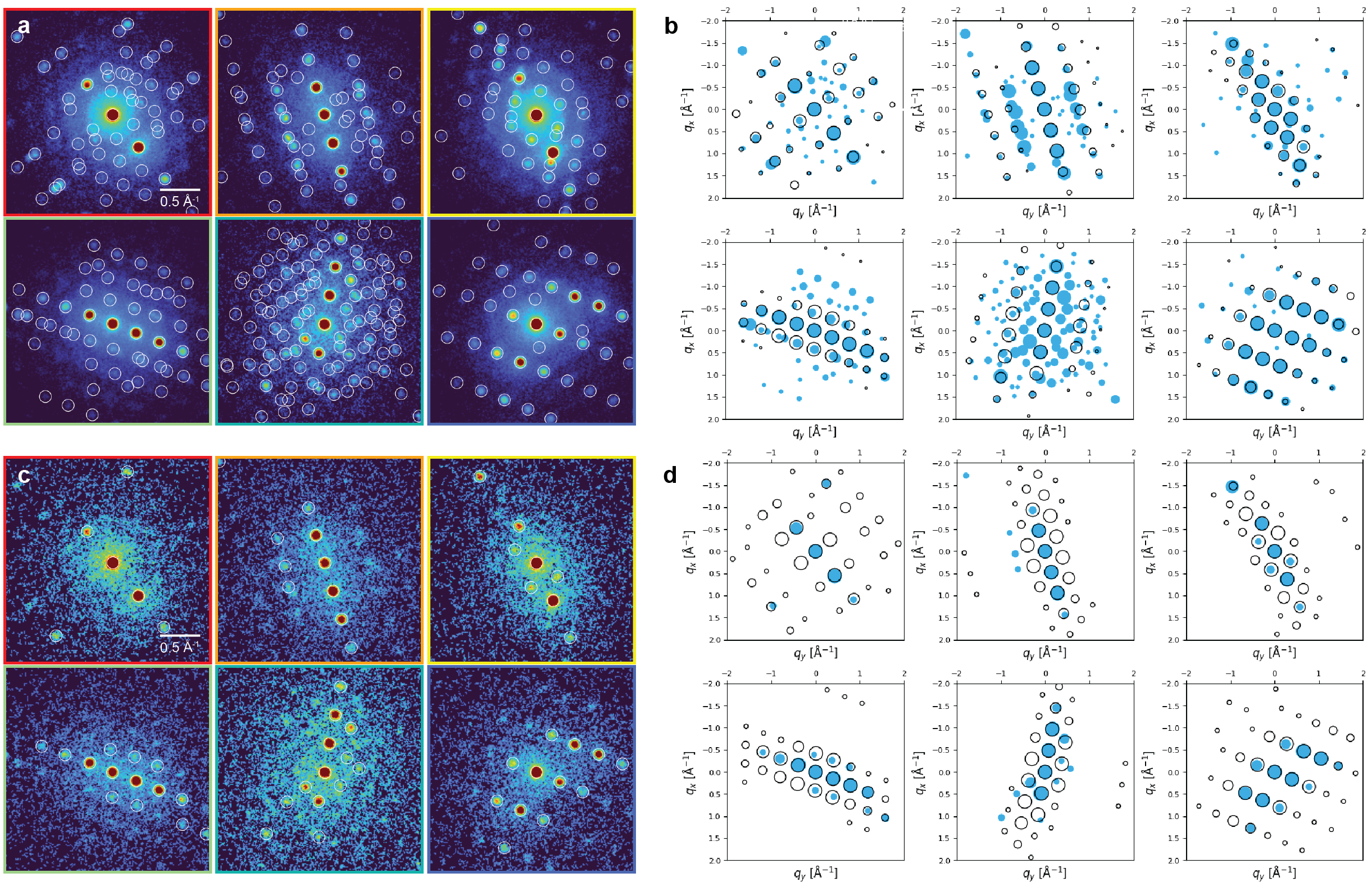}
  \caption{Bragg disk detection and orientation mapping on diffraction patterns after preprocessing but without clustering (a,b) and on raw data without both preprocessing and clustering (c,d). }
  \label{figS3}
\end{figure}

\clearpage

\FloatBarrier
\begin{figure}[!t]
  \centering
  \includegraphics[width=\linewidth]{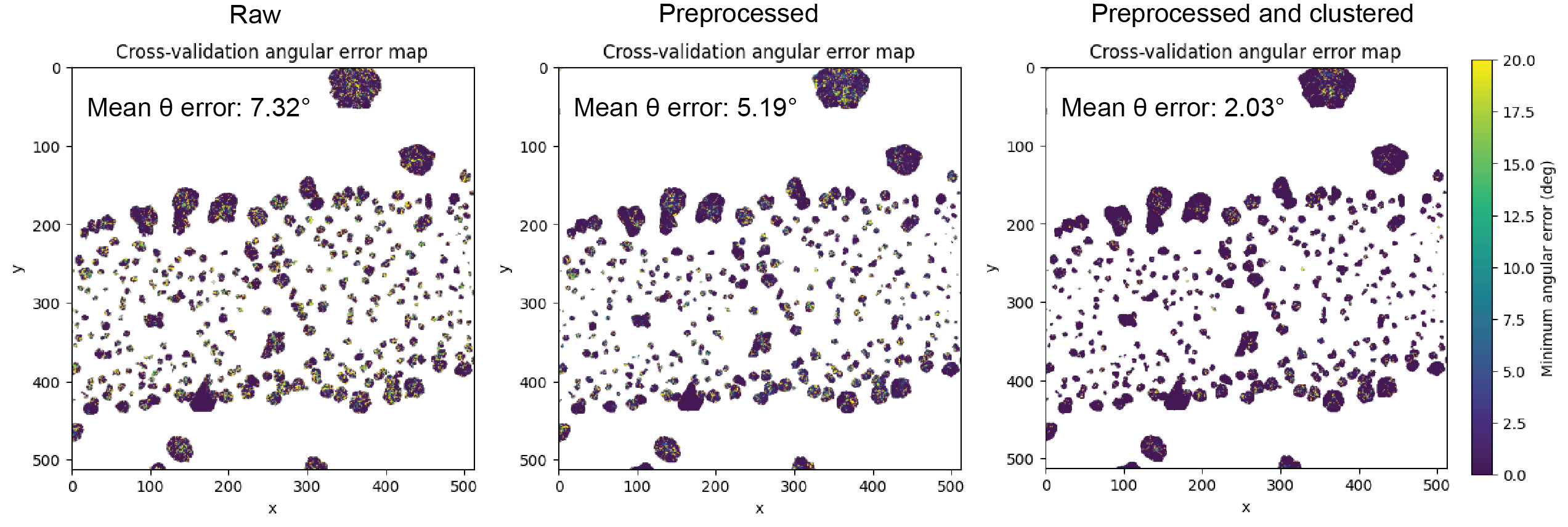}
  \caption{ Cross-validation angular error map of raw, preprocessed, and preprocessed and clustered dataset. }
  \label{figS4}
\end{figure}

\clearpage

\end{document}